\documentstyle[prl,cite,aps,multicol]{revtex}

\draft
\begin{document}

\renewcommand{\narrowtext}{\begin{multicols}{2}
\global\columnwidth20.5pc}
\renewcommand{\widetext}{\end{multicols} \global\columnwidth42.5pc}
\multicolsep = 8pt plus 4pt minus 3pt

\title{Resummation of QED Perturbation Series by Sequence
Transformations\\ and the Prediction of Perturbative Coefficients}

\author{U.~D.~Jentschura$^{1,}$\cite{InternetUJ},
J.~Becher$^{1}$,
E.~J.~Weniger$^{2,3}$,
and G.~Soff$^{1}$}

\address{$^{1}$Institut f\"{u}r Theoretische Physik,
TU Dresden, D-01062 Dresden, Germany\\
$^{2)}${\it Max-Planck-Institut~f\"{u}r Physik komplexer Systeme,
N\"{o}thnitzer Stra\ss{}e 38--40, 01087 Dresden, Germany}\\
$^{3}$Institut f\"{u}r Physikalische und
Theoretische Chemie, Universit\"{a}t
Regensburg, D-93040 Regensburg, Germany}

\maketitle

\begin{abstract}
We propose a method for the resummation of divergent perturbative
expansions in quantum electrodynamics and related field theories. The
method is based on a nonlinear sequence transformation and uses as
input data only the numerical values of a finite number of
perturbative coefficients. The results obtained in this way are for
alternating series superior to those obtained using Pad\'{e}
approximants. The nonlinear sequence transformation fulfills an
accuracy-through-order relation and can be used to predict
perturbative coefficients. In many cases, these predictions are closer
to available analytic results than predictions obtained using the
Pad\'{e} method.
\end{abstract}

\noindent

\narrowtext

%
% Introduction
%
\section{Introduction}
\label{Introduction}

Perturbation theory leads to the expansion of a physical quantity
${\cal P} (g)$ in powers of the coupling $g$,
\begin{equation}
\label{DefP}
{\cal P} (g) \; \sim \; \sum_{n=0}^{\infty} c_n \, g^n\, .
\end{equation}
The natural question arises as to how the power series on the
right-hand side is related to the (necessarily finite) quantity on the
left. It was pointed out in~\cite{Si1972} that perturbation theory is
unlikely to converge in any Lagrangian field theory. Generically, the
asymptotic behavior of the perturbative coefficients is assumed to be
of the form~\cite{VaZa1994}
\begin{equation}
\label{cAsymp}
c_n \sim K\,n^{\gamma}\,\frac{n!}{S^n} \, , \qquad n \to \infty \, ,
\end{equation}
where $K$, $\gamma$ and $S$ are constants. $S$ is related to the
first coefficient of the $\beta$ function of the underlying theory.

In view of the probable divergence of perturbation expansions in
higher order, a number of prescriptions have been proposed both for
the resummation of divergent perturbation series and for the
prediction of higher-order perturbative coefficients. A very important
method is the Borel summation procedure whose application to QED
perturbation series is discussed in~\cite{Og1956,DuHa1999}. The Borel
method, while being useful for the resummation of divergent series,
cannot be used for the prediction of higher-order perturbative
coefficients in an obvious way.

In recent years, Pad\'{e} approximants have become the standard tool to
overcome problems with slowly convergent and divergent power
series~\cite{BaGr1996}. Pad\'{e} approximants have also been used for
the prediction of unknown perturbative coefficients in quantum field
theory~\cite{SaLiSt1993,SaLiSt1995,SaElKa1995}. The $[ l / m ]$ Pad\'e
approximant to the quantity ${\cal P}(g)$ represented by the power
series~(\ref{DefP}) is the ratio of two polynomials $P_l(g)$ and
$Q_m(g)$ of degree $l$ and $m$, respectively,
\[
[ l / m ]_{\cal P}(g) \;=\; \frac{P_l(g)}{Q_m(g)}\;=\;
\frac{p_0 + p_1 \, g + \ldots + p_{l} \,  g^{l}}
{1 + q_1 \, g + \ldots + q_{m} \, g^{m}} \, .
\]
The polynomials $P_l(g)$ and $Q_m(g)$ are constructed so that the Taylor
expansion of the Pad\'e approximation agrees with the original input
series Eq.~(\ref{DefP}) up to terms of order $l+m$ in $g$,
\begin{equation}
\label{PadeOTA}
{\cal P}(g) \;-\; [ l / m ]_{\cal P}(g) \; = \; {\rm O}(g^{l+m+1})\, ,
\qquad g \to 0 \, .
\end{equation}
For the recursive computation of Pad\'{e} approximants we use Wynn's
epsilon algorithm~\cite{Wy1956a}, which in the case of the power
series~(\ref{DefP}) produces Pad\'{e} approximants according to
$\epsilon_{2k}^{(n)} = [ n+k / k ]_{\cal P} (g)$. Further details can be
found in Ch.~4 of~\cite{We1989}.

In this Letter, we advocate a different resummation scheme. For an
infinite series whose partial sums are $s_n = \sum_{j=0}^{n} \, a_j$,
the nonlinear (Weniger) sequence transformation with initial
element $s_0$ is defined as
[see~Eq.~(8.4-4) of~\cite{We1989}]:
\begin{equation}
\label{dWenTr}
{\delta}_n^{(0)} (\beta, s_0) \; = \; 
\frac {\displaystyle
\sum_{j=0}^{n} \; (- 1)^{j} \; {{n} \choose {j}} \;
\frac {(\beta + j)_{n-1}} {(\beta + n)_{n-1}} \;
\frac {s_{j}} {a_{j + 1}} }
{\displaystyle
\sum_{j=0}^{n} \; (- 1)^{j} \; {{n} \choose {j}} \;
\frac {(\beta + j)_{n-1}} {(\beta + n)_{n-1}} \;
\frac {1} {a_{j + 1}} } \, ,
\end{equation}
where $(a)_m = \Gamma(a + m)/\Gamma(a)$ is a Pochhammer symbol.
The shift parameter $\beta$ is usually chosen as $\beta = 1$, and this
choice will be exclusively used here (see also~\cite{We1989}). The
power of the $\delta$ transformation and related transformations
[e.g., the Levin transformation, Eq.~(7.3-9) of~\cite{We1989}] is due
to the fact that explicit estimates for the truncation error of the
series are incorporated into the convergence acceleration or
resummation process (see Ch.~8 of~\cite{We1989}). Note that the
$\delta$ transformation~(\ref{dWenTr}) has lead to numerically stable
and remarkably accurate results~\cite{WeCiVi1993,We1996d} in the
resummation of the perturbative series of the quartic, sextic and
octic anharmonic oscillator whose coefficients display a similar
factorial pattern of divergence as the quantum field theoretic
coefficients indicated in Eq.~(\ref{cAsymp}).

\widetext
%
% Table 1
%
\begin{table}[tbh]
\caption{Resummation of the perturbation series (\ref{Bperser}) for
$g_{\rm B}=10$. Results are given in terms of the dimensionless
function ${\bar S}_{\rm B} = 10^2 \, \bigl[(8 \pi^2)/(-e^2 B^2 g_{\rm
B}) \bigr] \, S_{\rm B}$.  Apparent convergence is indicated by
underlining.}
\label{table1}
%
%\squeezetable
\begin{tabular}{rrrrr}
\multicolumn{1}{c}{$n$}%
& \multicolumn{1}{c}{$s_n$}%
& \multicolumn{1}{c}
{$\bigl[ {[\mkern - 2.5 mu [(n+1)/2] \mkern - 2.5 mu ]} / 
{[\mkern - 2.5 mu [n/2] \mkern - 2.5 mu ]} \bigr]$}%
& \multicolumn{1}{c}
{$d_{n-1}^{(0)} \bigl(1, s_0 (g_{\rm B}) \bigr)$ }%
& \multicolumn{1}{c}
{$\delta_{n-1}^{(0)} \bigl(1, s_0 (g_{\rm B}) \bigr)$ 
\rule[-6pt]{0pt}{4\jot} } \\
\hline
$ 1 $&
  $10.476$&
  $10.476~190~476$&
  $-2.222~222~222$&
  $-2.222~222~222$\\
$ 2 $&
  $-243.492$&
  $\underline{}-1.617~535~903$&
  $-1.617~535~903$&
  $-1.617~535~903$\\
$ 3 $&
  $10~530.918$&
  $\underline{}4.627~654~271$&
  $\underline{-0}.820~833~551$&
  $\underline{-0}.820~833~551$\\
$ 4 $&
  $-774~888.106$&
  $\underline{}-1.401~288~801$&
  $\underline{-0}.588~575~814$&
  $\underline{-0}.659~817~926$\\
$ 5 $&
  $8.674~647 \times 10^{7}$&
  $\underline{}2.773~159~300$&
  $\underline{-0.8}64~617~071$&
  $\underline{-0.7}33~843~307$\\
$\dots$ &
  $\dots$ &
  $\dots$ &
  $\dots$ &
  $\dots$ \\
$ 60 $&
  $-3.652~544 \times 10^{201}$&
  $\underline{-0.9}20~487~125$&
  $5.992~187 \times 10^{12}$&
  $\underline{-0.805~633~9}81$\\
$ 61 $&
  $5.553~434 \times 10^{205}$&
  $\underline{-0}.400~319~939$&
  $1.385~114 \times 10^{13}$&
  $\underline{-0.805~633~9}80$\\
$ 62 $&
  $-8.721~566 \times 10^{209}$&
  $\underline{-0.9}18~054~104$&
  $-4.131~495 \times 10^{13}$&
  $\underline{-0.805~633~97}9$\\
$ 63 $&
  $1.414~066 \times 10^{214}$&
  $\underline{-0}.411~140~364$&
  $-8.500~694 \times 10^{13}$&
  $\underline{-0.805~633~97}8$\\
$ 64 $&
  $-2.365~759 \times 10^{218}$&
  $\underline{-0.9}15~746~814$&
  $2.890~004 \times 10^{14}$&
  $\underline{-0.805~633~97}7$\\
$ 65 $&
  $4.082~125 \times 10^{222}$&
  $\underline{-0}.421~331~007$&
  $5.272~267 \times 10^{14}$&
  $\underline{-0.805~633~97}6$\\
$ 66 $&
  $-7.261~275 \times 10^{226}$&
  $\underline{-0.9}13~555~178$&
  $-2.050~491 \times 10^{15}$&
  $\underline{-0.805~633~975}$\\
$ 67 $&
  $1.330~921 \times 10^{231}$&
  $\underline{-0}.430~946~630$&
  $-3.296~170 \times 10^{15}$&
  $\underline{-0.805~633~975}$\\
$\dots$ &
  $\dots$ &
  $\dots$ &
  $\dots$ &
  $\dots$ \\
\hline
exact &
  ${-0.805~633~975}$&
  ${-0.805~633~975}$&
  ${-0.805~633~975}$&
  ${-0.805~633~975}$\\
\end{tabular}
\end{table}
\narrowtext

We consider as a model problem the QED effective action in the
presence of a constant background magnetic field for which the exact
nonperturbative result can be expressed as a proper-time integral:
\begin{equation}
\label{SRB}
S_{\rm B} = - \frac{e^2 B^2}{8 \pi^2} \!
\int\limits_0^\infty \frac{{\rm d}s}{s^2} \!
\left\{\coth s \! - \! \frac{1}{s} \! - \! \frac{s}{3} \right\}
\exp\!\left(-\frac{m_{\rm e}^2}{e\,B} s\right)\,.
\end{equation}
Here, $B$ is the magnetic field strength, and $e$ is the elementary
charge. The general result for arbitrary $E$ and $B$ field can be
found in Eq.~(3.49) in~\cite{Sc1951} and in Eq.~(4-123)
in~\cite{ItZu1980}. The nonperturbative result for $S_{\rm B}$ can be
expanded in powers of the effective coupling $g_{\rm B} = e^2 B^2 /
m_{\rm e}^4$, which results in the divergent asymptotic series
\begin{eqnarray}
\label{Bperser}
S_{\rm B} & \sim &
- \frac{2 e^2 B^2 }{\pi^2} \, g_{\rm B} \,
\sum_{n=0}^\infty \; c_n \; g_{\rm B}^n \, , \qquad g_{\rm B} \to 0 \, .
\end{eqnarray}
The expansion coefficients
\begin{equation}
\label{cn}
c_n = \frac{(-1)^{n + 1} \; 4^n \;
\left|{\cal B}_{2n+4}\right| }{(2n+4)(2n+3)(2n+2)}\,,
\end{equation}
where ${\cal B}_{2n+4}$ is a Bernoulli number, display an alternating
sign pattern and grow factorially in absolute magnitude,
\begin{equation}
\label{cnasymptotic}
c_n \; \sim \; \frac{(-1)^{n+1}}{8} \; \frac{\Gamma(2n+2)}{\pi^{2n+4}}
\left( 1 + {\rm O}({2^{-(2n+4)}}) \right)
\end{equation}
as $n \to \infty$. The series differs from ``usual'' perturbation
series in quantum field theory by the distinctive property that 
all perturbation theory coefficients are known. 

The numerical results in the fifth column of Table~\ref{table1} show
that the application of the $\delta$ transformation~(\ref{dWenTr}) to
the partial sums $s_n (g_{\rm B})$ of the perturbation
series~(\ref{Bperser}) produces convergent results even for a coupling
constant as large as $g_{\rm B} = 10$. In the third column of
Table~\ref{table1}, we display
the sequence
\begin{displaymath}
[0/0], [1/0], [1/1], \ldots, [\nu/\nu], [\nu+1/\nu], [\nu+1/\nu+1], 
\ldots
\end{displaymath}
of Pad\'{e} approximants, which were computed using Wynn's epsilon
algorithm~\cite{Wy1956a}. With the help of the notation ${[\mkern -
2.5 mu [x] \mkern - 2.5 mu ]}$ for the integral part of $x$, the
elements of this sequence of Pad\'{e} approximants can be written
compactly as $\bigl[ {[\mkern - 2.5 mu [(n+1)/2] \mkern - 2.5 mu ]} /
{[\mkern - 2.5 mu [n/2] \mkern - 2.5 mu ]} \bigr]$. Obviously,
Pad\'{e} approximants converge too slowly to the exact result to be
numerically useful. The Levin $d$ transformation defined in
Eq.~(7.3-9) in~\cite{We1989}, which is included because it is closely
related to the $\delta$ transformation~(\ref{dWenTr}), fails to
accomplish a resummation of the perturbation series, as shown in the
fourth column of Table~\ref{table1}.

So far, predictions for unknown perturbative coefficients were usually
obtained using Pad\'{e} approximants. The accuracy-through-order
relation~(\ref{PadeOTA}) implies that the Taylor expansion of a Pad\'{e}
approximant reproduces all terms used for its construction. The next
coefficient obtained in this way is usually interpreted as the
prediction for the first unknown series coefficient (see,
e.g.,~\cite{SaLiSt1993,SaLiSt1995,SaElKa1995}). The
$\delta$
transformation~(\ref{dWenTr}), when applied to the partial sums ${\cal
P}_n (g)$ of the power series~(\ref{DefP}), fulfills the
accuracy-through-order relation~\cite{WeCiVi1993}:
\begin{equation}
\label{dWenOTA}
{\cal P}(g) \, - \, \delta_n^{(0)} \bigr(1,{\cal P}_0 (g) \bigl) 
\; = \; {\rm O}(g^{n+2}) \, , \qquad g \to 0 \, .
\end{equation}
Upon re-expansion of the $\delta$ transform a prediction for the next
higher-order term in the perturbation series may therefore be
obtained. 

In Table~\ref{table2} we compare predictions for the coefficients
$c_n$ of the perturbation series~(\ref{Bperser}) obtained by
re-expanding the Pad\'{e} approximants $\bigl[ {[\mkern - 2.5 mu [n/2]
\mkern - 2.5 mu ]} / {[\mkern - 2.5 mu [(n-1)/2] \mkern - 2.5 mu ]}
\bigr]$ and the transforms $\delta_{n-2}^{(0)} \bigl(1, s_0 (g_{\rm
B}) \bigr)$, which were computed from the partial sums $s_{0} (g_{\rm
B})$, $s_{1} (g_{\rm B})$, \ldots, $s_{n-1} (g_{\rm B})$. For higher
orders of perturbation theory in particular, the Weniger
transformation yields clearly the best results, whereas for low orders
the improvement over Pad\'{e} predictions is only gradual. For
example, let us assume that for a particular problem only three
coefficients $c_0$, $c_1$ and $c_2$ are available and $c_3$ should be
estimated by a rational approximant. Because of the accidental
equality $[1/1]_{\cal P}(g) = \delta^{(0)}_1\bigl(1,{\cal
P}_0(g)\bigr)$, the predictions for $c_3$ obtained using the Pad\'{e}
scheme and the $\delta$ transformation, are equal. Differences between
the Pad\'{e} predictions and those obtained using the $\delta$
transformation start to accumulate in higher order.

\widetext
%
% Table 2
%
\begin{table}[tbh]
\caption{Prediction of perturbative coefficients for the power
series~(\ref{Bperser}). Results are given for the scaled dimensionless
power series $S'_{\rm B} = \bigl[(8 \pi^2)/(-e^2 B^2 g_{\rm B}) \bigr]
\, S_{\rm B}$.  First column: order of perturbation theory. Second
column: exact coefficients. Third and fourth column: predictions
obtained by re-expanding Pad\'{e} approximants and Weniger transforms,
respectively.}
\label{table2}
%
%\squeezetable
\begin{tabular}{rlll}
\multicolumn{1}{c}{$n$}%
& \multicolumn{1}{c}{exact}%
& \multicolumn{1}{c}
{$\bigl[ {[\mkern - 2.5 mu [n/2] \mkern - 2.5 mu ]} / 
{[\mkern - 2.5 mu [(n-1)/2] \mkern - 2.5 mu ]} \bigr]$}%
& \multicolumn{1}{c}
{$\delta_{n-2}^{(0)} \bigl(1, {s}_0 (g_{\rm B}) \bigr)$ 
\rule[-6pt]{0pt}{4\jot} } \\
\hline
$ 3 $&
  $+0.107~744~107$&
  $+0.050~793~650$&
  $+0.050~793~650$\\
$ 4 $&
  $-0.785~419~025$&
  $-0.457~096~214$&
  $-0.537~632~214$\\
$\dots$ &
  $\dots$ &
  $\dots$ &
  $\dots$ \\
$ 14 $&
  ${-2.181~588~772 \times 10^{15}}$&
  $\underline{-2.1}70~458~614 \times 10^{15}$&
  $\underline{-2.181~5}74~607 \times 10^{15}$\\
$ 15 $&
  ${+2.055~682~756 \times 10^{17}}$&
  $\underline{+2.0}49~236~087 \times 10^{17}$&
  $\underline{+2.055~6}78~921 \times 10^{17}$\\
$ 16 $&
  ${-2.199~481~257 \times 10^{19}}$&
  $\underline{-2.19}4~962~521 \times 10^{19}$&
  $\underline{-2.199~48}0~091 \times 10^{19}$\\
$\dots$ &
  $\dots$ &
  $\dots$ &
  $\dots$ \\
$ 24 $&
  ${-1.711~360~421 \times 10^{37}}$&
  $\underline{-1.711}~272~235 \times 10^{37}$&
  $\underline{-1.711~360~421} \times 10^{37}$\\
$ 25 $&
  ${+4.421~625~118 \times 10^{39}}$&
  $\underline{+4.421}~484~513 \times 10^{39}$&
  $\underline{+4.421~625~118} \times 10^{39}$\\
$ 26 $&
  ${-1.234~699~825 \times 10^{42}}$&
  $\underline{-1.234~6}74~716 \times 10^{42}$&
  $\underline{-1.234~699~825} \times 10^{42}$\\
$\dots$ &
  $\dots$ &
  $\dots$ &
  $\dots$ \\
\end{tabular}
\end{table}
\narrowtext

We now turn to the case of the uniform background electric field, for
which the effective action reads~\cite{Sc1951}
\[
S_{\rm E} = \frac{e^2 E^2}{8 \pi^2} \!
\int\limits_0^\infty \! \! \frac{{\rm d}s}{s^2} \!
\left\{\coth s \! - \! \frac{1}{s} \! - \! \frac{s}{3} \right\}
\exp\left[{\rm i} \!
\left(\frac{m_{\rm e}^2}{e\,E} \! + \!
{\rm i}\,\epsilon\right)\,s\right]\,.
\]
This result can be derived from~(\ref{SRB}) by the replacements $B \to
{\rm i}\,E$ and the inclusion of the converging factor. With the
convention $g_{\rm E} = e^2\,E^2/m_{\rm e}^4$ the divergent asymptotic
series
\begin{eqnarray}
\label{Eperser}
S_{\rm E} & \sim &
\frac{2 e^2 E^2 }{\pi^2} \, g_{\rm E} \,
\sum_{n=0}^\infty \; c'_n \; g_{\rm E}^n \, , \qquad g_{\rm E} \to 0 \, ,
\end{eqnarray}
is obtained. The expansion coefficients
\begin{equation}
\label{cnprime}
c'_n = \frac{ 4^n \; \left|{\cal B}_{2n+4}\right| }{(2n+4)(2n+3)(2n+2)}
\end{equation}
display a nonalternating sign pattern, but are equal in magnitude to the
magnetic field case [cf.~Eq.~(\ref{cn})]. For physical values of $g_{\rm
E}$, i.e., for $g_{\rm E} > 0$, there is a cut in the complex plane, and
the nonvanishing imaginary part for $S_{\rm E}$ gives the
pair-production rate. As is well known, resummation procedures for
(nonalternating) divergent series usually fail when the coupling $g$
assumes values on the cut in the complex plane~\cite{We1989}. The Borel
method fails because of the poles on the integration contour in the
Borel integral~\cite{DuHa1999}. The $\delta$ transformation and Pad\'{e}
approximations fail for reasons discussed in~\cite{We1989}
and~\cite{Pi1999}, respectively.

We now come to an important observation which to the best of our
knowledge has not yet been addressed in the literature: the
prediction of perturbative coefficients by nonlinear sequence
transformations may even work if the resummation of the divergent
series fails, i.e.~if the coupling $g$ lies on the cut. A general
divergent series whose coefficients are nonalternating in sign,
evaluated for positive coupling, corresponds to a series with
alternating coefficients, evaluated for negative
coupling. Alternating series can be resummed with the $\delta$
transformation in many cases, and predictions for higher-order
coefficients should therefore be possible for both the alternating
and the nonalternating case. For example, the perturbative
coefficients in Eqs.~(\ref{cn}) and~(\ref{cnprime}) differ only in the
sign pattern, not in their magnitude. As shown in Table~\ref{table3},
rational approximants to the series (\ref{Bperser}) and
(\ref{Eperser}) produce, after the re-expansion in the coupling, the
same predictions up to the different sign pattern.

We would like to stress here that the resummation procedure and the
prediction scheme presented in this Letter also work for higher-order
terms in the derivative expansion of the QED effective
action~\cite{BeJeMHMoWeSo1999}. The resummation also works for the
partition function for the zero-dimensional $\phi^4$ theory which is
discussed in~\cite{ItZu1980} (p.~464) and is used in~\cite{ZJ1996} as
a paradigmatic example for the divergence of perturbative expansions
in quantum field theory. Results will be presented in detail
elsewhere~\cite{BeJeMHMoWeSo1999}.

An interesting and more ``realistic'' application is given by the
$\beta$ function of the Higgs boson coupling in the standard
electroweak model~\cite{DuJa1998}. In the $\overline{\rm MS}$
renormalization scheme, five coefficients of this $\beta$ function are
known. Using the first four coefficients, the ``prediction'' for the
fifth coefficient (which is known) may be obtained and compared to the
analytic result. Using the transformation $\delta^{(0)}_2$ a
prediction of $\beta_4 \approx 4.404 \times 10^7$ is obtained which is
closer to the analytic result of $\beta_4 \approx 4.913 \times
10^7$ than the predictions obtained using the $[2/1]$ and $[1/2]$
Pad\'{e} approximants (these yield $\beta_4 \approx 3.969 \times 10^7$
and $\beta_4 \approx 4.188 \times 10^7$, respectively). The prediction
for the unknown coefficient $\beta_5$ obtained using $\delta^{(0)}_3$
is $\beta_5 \approx -3.938 \times 10^9$ as compared to $\beta_5
\approx -3.756 \times 10^9$ from the $[2/2]$ Pad\'{e} approximant.

\begin{center}
\begin{minipage}{8.0cm}
%
% Table 3
%
\begin{table}[htb]
\caption{Prediction of perturbative coefficients $c'_n$ for the {\em
electric} background field~(\ref{Eperser}). Results are given for the
scaled dimensionless power series $S'_{\rm E} = \bigl[ (8 \pi^2)/(e^2
\, E^2 g_{\rm E}) \bigr] S_{\rm E}$.}
\label{table3}
\begin{tabular}{rll}
\multicolumn{1}{c}{$n$}
& \multicolumn{1}{c}{exact}
& \multicolumn{1}{c}
{${\delta}_{n-2}^{(0)} \bigl(1, s_0 (g_{\rm E}) \bigr)$} \\
\hline        
  $\dots$ &
  $\dots$ &
  $\dots$ \\
$ 14 $&
  ${2.181~588 \times 10^{15}}$&
  $\underline{2.181~5}74 \times 10^{15}$\\
$ 15 $&
  ${2.055~682 \times 10^{17}}$&
  $\underline{2.055~6}78 \times 10^{17}$\\
$ 16 $&
  ${2.199~481 \times 10^{19}}$&
  $\underline{2.199~48}0 \times 10^{19}$\\
$\dots$ &
  $\dots$ &
  $\dots$ \\
$ 24 $&
  ${1.711~360 \times 10^{37}}$&
  $\underline{1.711~360} \times 10^{37}$\\
$ 25 $&
  ${4.421~625 \times 10^{39}}$&
  $\underline{4.421~625} \times 10^{39}$\\
$ 26 $&
  ${1.234~699 \times 10^{42}}$&
  $\underline{1.234~699} \times 10^{42}$\\
$\dots$ &
  $\dots$ &
  $\dots$ \\
\end{tabular}
\end{table}
\end{minipage}
\end{center}

For the $\beta$ function of the scalar $\phi^4$ theory the situation
is similar to the Higgs boson case. Five coefficients are known
analytically~\cite{KlEtAl1991}.  Again, the prediction for the fifth
coefficient obtained using the transformation $\delta^{(0)}_2$
($1251.3$) is closer to the analytic result of $1424.3$ than the
predictions from the $[2/1]$ and $[1/2]$ Pad\'{e} approximants which
yield values of $1133.5$ and $1187.5$, respectively.  For the unknown
sixth coefficient, a prediction of $- 1.70 \times 10^4$ is obtained
using $\delta^{(0)}_3$ whereas the $[2/2]$ Pad\'{e} approximant yields
$- 1.63 \times 10^4$.

We have shown that the $\delta$ transformation (\ref{dWenTr}) can be
used to accomplish a resummation of alternating divergent
perturbation series whose coefficients diverge factorially. In many
cases, the $\delta$ transforms converge faster to the nonperturbative
result than Pad\'{e} approximants.  The $\delta$ transformation uses
as input data only the numerical values of a finite number
of perturbative coefficients.  We stress here that the factorial
divergence is expected of general perturbative expansions in
quantum field theory [see Eq.~(\ref{cAsymp})]. The Weniger $\delta$
transformation can be used for the prediction of higher-order
coefficients of alternating and nonalternating factorially
divergent perturbation series. Both in model problems and in more
realistic applications, the $\delta$ transformation yields improved
predictions (compared to Pad\'{e} approximants).  It appears that the
potential of sequence transformations, notably the $\delta$
transformation, has not yet been widely noticed in the field of
large-order perturbation theory.

U.D.J. acknowledges helpful conversations with M. Meyer-Hermann,
P. J. Mohr and K. Pachucki. E.J.W. acknowledges support from the
Fonds der Chemischen Industrie. G.S. acknowledges continued support
from BMBF, GSI and DFG.

\widetext

\end{document}